\documentclass[12pt]{article}

\usepackage[dvipsnames]{xcolor}
\usepackage{amsmath}
\usepackage{amssymb}

\newtheorem{assumption}{Assumption}
\newtheorem{definition}{Definition}
\newtheorem{example}{Example}
\usepackage{caption}
\usepackage{dsfont}
\usepackage{tikz}
\usepackage[english]{babel}
\usepackage[utf8]{inputenc}
\usepackage{times}
\usepackage{graphics}
\usepackage{graphicx}
\usepackage[T1]{fontenc}
\usepackage[official,right]{eurosym}
\usepackage{url}
\usepackage{eso-pic}

\title{Subjective fairness in algorithmic decision-support}
\author{Sarra Tajouri, Alexis Tsoukiàs \\ CNRS-LAMSADE, PSL, Université Paris Dauphine}
\date{}

\begin{document}

\thispagestyle{empty}

\enlargethispage*{8cm}
 \vspace*{-38mm}

\AddToShipoutPictureBG*{\includegraphics[width=\paperwidth,height=\paperheight]{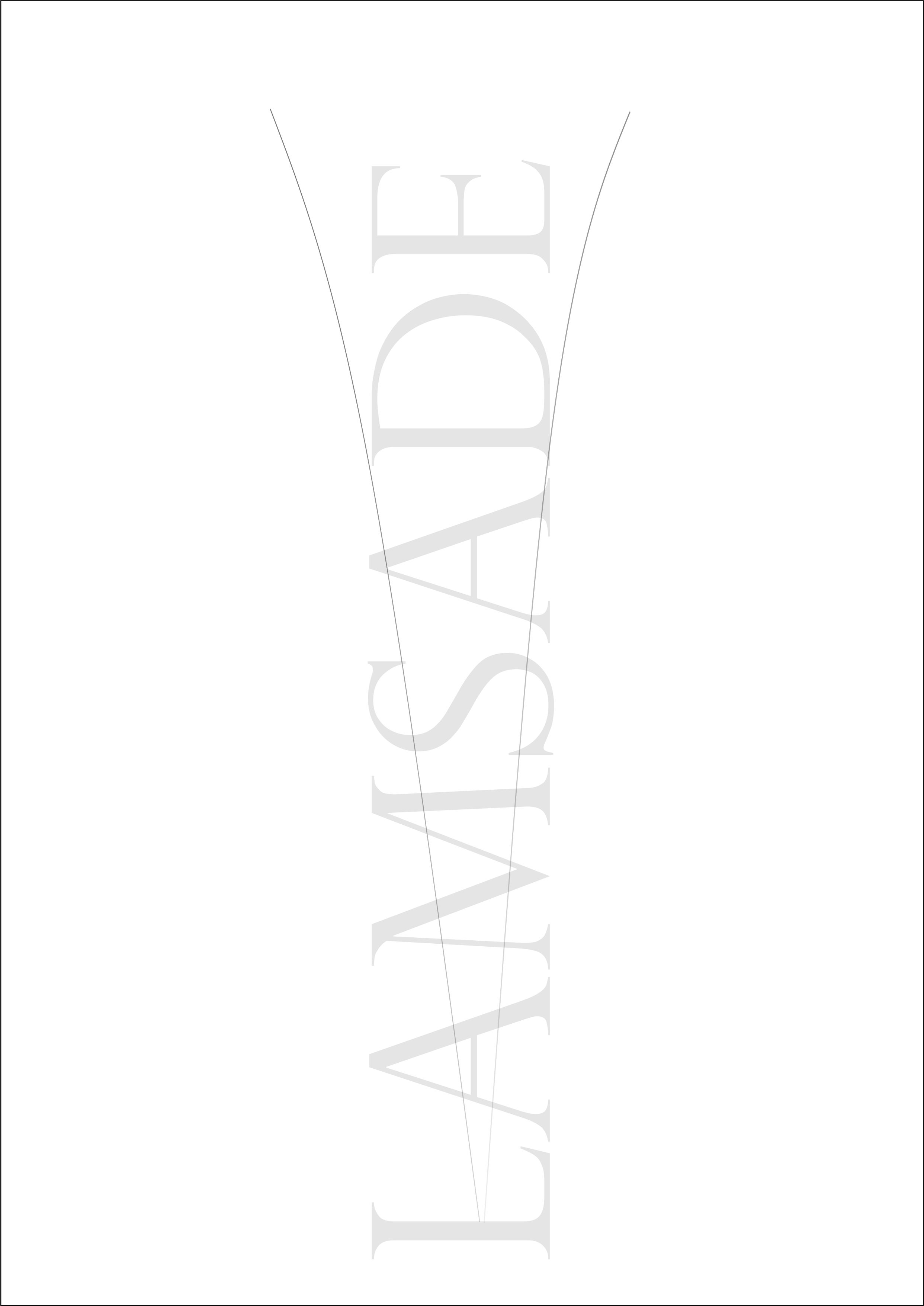}}

\begin{minipage}{24cm}
 \hspace*{-28mm}
\begin{picture}(500,700)\thicklines
 \put(60,670){\makebox(0,0){\scalebox{0.7}{\includegraphics{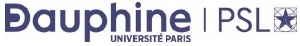}}}}
 \put(60,70){\makebox(0,0){\scalebox{0.3}{\includegraphics{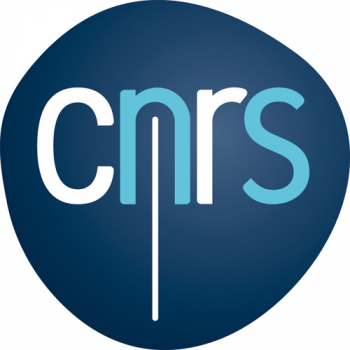}}}}
 \put(320,350){\makebox(0,0){\Huge{CAHIER DU \textcolor{BurntOrange}{LAMSADE}}}}
 \put(140,10){\textcolor{BurntOrange}{\line(0,1){680}}}
 \put(190,330){\line(1,0){263}}
 \put(320,310){\makebox(0,0){\Huge{\emph{409}}}}%XXX number of the Cahier: provided by Eleni
 \put(320,290){\makebox(0,0){June 2024}}
 \put(320,210){\makebox(0,0){\Large{Subjective fairness in algorithmic decision-support}}}
 %\put(320,190){\makebox(0,0){\Large{subtitle}}}
 \put(320,100){\makebox(0,0){\Large{Sarra Tajouri, Alexis Tsoukiàs}}}
 \put(320,670){\makebox(0,0){\Large{\emph{Laboratoire d'Analyse et Mod\'elisation}}}}
 \put(320,650){\makebox(0,0){\Large{\emph{de Syst\`emes pour l'Aide \`a la D\'ecision}}}}
 \put(320,630){\makebox(0,0){\Large{\emph{UMR 7243}}}}
\end{picture}
\end{minipage}

\newpage

\addtocounter{page}{-1}

\maketitle

\abstract{
The treatment of fairness in the decision-making literature usually consist in quantifying fairness through objective measures. This work takes a critical stance to highlight the limitations of these approaches (group fairness and individual fairness) using sociological insights.
    First, we expose how these metrics often fail to reflect societal realities. By neglecting crucial historical, cultural, and social factors, they fall short of capturing all discriminatory practices.
    Second, we redefine fairness as a subjective property moving from a top-down to a bottom-up approach. This shift allows the inclusion of diverse stakeholders' perceptions, recognizing that fairness is not merely about objective metrics but also about individuals' views on their treatment.   
    Finally, we aim to use explanations as a mean to achieve fairness. Our approach employs explainable clustering to form groups based on individuals’ subjective perceptions to ensure that individuals who see themselves as similar receive similar treatment. We emphasize the role of explanations in achieving fairness, focusing not only on procedural fairness but also on providing subjective explanations to convince stakeholders of their fair treatment.

}

\newpage

\section{Introduction}
    The pursuit of fairness, closely linked to the discourse on justice, has long captivated the minds of philosophers \cite{sen1979equality, dworkin1981equality, cohen1989currency, arneson1991equality, rawls2001justice}, sociologists \cite{weber1978distribution, bourdieu1990reproduction, collins2019intersectionality}, and economists \cite{roemer1996theories, arrow1974discrimination}. 
    Notably, John Rawls' theory of justice as fairness, outlined in his seminal work "A Theory of Justice" \cite{rawls2001justice}, posits that societal arrangements should be structured to benefit the least advantaged members. Rawls argues for a hypothetical social contract, where individuals, behind a "veil of ignorance" about their own characteristics, would agree on principles of justice. He proposes two principles: the equal basic liberties for all and social and economic inequalities that are arranged to benefit the least advantaged, known as the \textit{Difference Principle}.
    This work does not focus on fairness purely as a philosophical concept. Instead, we propose a novel approach to fairness in algorithmic decision-making that integrates the perceptions of individuals who collectively agree on what constitutes fairness.

    Indeed, the pervasive use of algorithms in social and economic contexts has raised a significant concern. Many instances of algorithmic discrimination have been reported over the past decade including the investigation by \textit{ProPublica} which exposed racial biases in Northpointe's COMPAS tool, a recidivism risk assessment algorithm \cite{angwin2016machine}. These cases of algorithmic discrimination stem from models trained on either biased datasets \cite{danks2017algorithmic}, which lack diversity and fail to accurately represent society \cite{suresh2019framework}, or datasets that simply mirror the inequalities inherent in certain segments of society. Without adequate consideration for debiasing or corrective measures, these algorithms will perpetuate and exacerbate social injustices.  

    Consequently, fairness has become a fast-growing area of interest within the scientific community, giving rise to a proliferation of research papers \cite{dwork2012fairness, hardt2016equality, CorbettDaviesetal2017, makhlouf2020survey, mehrabi2021survey} that have crystallized into a distinct field in computer science. This rise is due to the recognition that fairness is intrinsically linked to the collective well-being of our society. Numerous works have sought to formalize fairness within a decision process using mathematical constraints or metrics \cite{hardt2016equality, zafar2017fairness, verma2018fairness}. Although those techniques are well founded and produce promising results, they often fall short in incorporating the social dimension of fairness. In parallel, extensive work in social sciences has focused on this aspect, offering valuable insights that can significantly enrich the interdisciplinary perspective on fairness in computer science.

    A decision process consists of multiple stages. When we discuss fairness, it is always with respect to a specific stage of the process and directed towards a particular stakeholder. As such, fair process refers to the impartiality of procedures and methods used in making a decision. Whereas fair recommendation (suggestions of a decision-support system) and fair decision (final result) concern the outcomes. This distinction will be discussed notably in section \ref{section2}.
    Following that, we expose some of the limitations of traditional fairness approaches, highlighting how they lack to align with societal realities, drawing on sociological perspectives. 
    Then, we propose a novel approach viewing fairness in the decision-making process through a subjective lens.
    Central to our perspective is the inclusion of those impacted by a decision in the assessment of fairness, empowering them to determine if they are treated fairly.
    Our methodology for achieving subjective fairness employs explanations as a tool to construct justifications, aiming to convince stakeholders of the fairness in the process.

\section{Decision framework}
\label{section2}
    \subsection{Decision, process and recommendation}
    \label{section2_1}
    Decision-making has been studied in several disciplines ranging from cognitive studies \cite{burke2005improving,frydman2016psychology} to management science \cite{daellenbach2017management} and from economy to organizational studies \cite{simon2013administrative}. From an abstract mathematical perspective, any partitioning of set according to a subject's preferences is a decision problem \cite{colorni2024decision}. In economics, a decision is an irreversible allocation of resources to individuals for an objective achievement such as maximizing utility, profits etc..

    Contrary to viewing decisions as a single act, it should be regarded as a process. Simon \cite{simon2013administrative} defined a decision process (DP) as a series of means and ends connected in a hierarchical chain to achieve an objective including predictions about decision behavior. According to the classical theory, accurate prediction of behavior can be achieved by considering the environment together with the strong assumption of \textit{perfect rationality}. However, when dealing with imperfect competition and decision-making under uncertainty, such as not knowing all possible alternatives or external events, models of \textit{bounded rationality}, subjectively defined and only valid within specific contexts, are more suitable to provide more realistic explanations of human decision-making behavior \cite{simon1979rational}. 

    Decision-aiding process (DAP) is one form of decision process that involves more than one stakeholder. To simplify we can consider that it includes a decision-maker, who has domain knowledge concerning the decision process, and an analyst, who has technical expertise. The objective is to reach a consensus between these two actors that responds to the initial problem of the decision-maker \cite{tsoukias2008decision}.  
    
    Then, the increasing availability of data and the expansion of computers capacity led to the rise of using automated decision making systems (ADMS) where algorithms could autonomously take decisions without human intervention \cite{tsoukias2008decision}.
    From a computer science perspective, decisions can be viewed as the output of an algorithmic process with binary outcomes. ADMS are particularly suited for frequent generic decisions where speed is critical, supported by available collected data. Credit scoring is one example of ADMS where the decision problem is standardized: assigning a risk score to each candidate. This process is repetitive and depends on the output of software developed by the analyst(s). The decision-maker then uses these scores to make final decisions on whether to grant credit or not.
    
    It's crucial to recognize that decisions carry responsibility and liability due to their potential unintended consequences  \cite{tsoukias2020social}. 
    But since the decisions are automated, the responsibility is often diffuse, shared among human stakeholders involved in the process design and implementation. 
    ADMS actually propose a recommendation to the decision-maker who takes the responsibility to decide afterwards. However, most of the time, humans may lack the capacity to thoroughly evaluate each recommendation, leading them to accept suggestions without scrutiny. Consequently, the control over the final decision doesn't guarantee control over the outcome, as ''recommendations are often treated as decisions''\cite{tsoukias2020social}. 

    These advancements led to a proliferation of literature about the implications of such automated decision-making, including discussions on AI ethics, fairness, accountability, transparency etc.. In fact, algorithmic decisions have unveiled inherent biases in society and issues of discrimination against minorities that were embedded in the data we used to automate decisions. However, it is important to keep in mind that while correcting algorithmic outcomes is necessary, it alone cannot address broader societal issues. The most effective response to social challenges remains rooted in human and political action rather than purely technological solutions.

    \subsection{Fairness of what ?}
    Practically, some of the decision processes concern algorithms that suggest a certain action to be undertaken in a high-stake context. It could concern the distribution of financial resources, economic or educational opportunities, granting bail etc..

    The EU AI act \cite{aiact}, the first regulation on artificial intelligence proposed by the European commission, attempts to regulate the use of AI systems using a risk-based approach, distinguishing unacceptable, high and low risk. 
    High-stake decisions fall under the definition advanced by the AI Act as "\textit{systemic risk at Union level means a risk that is specific to the high-impact capabilities [...], having a significant impact on the internal market due to its reach, and with actual or reasonably foreseeable negative effects on public health, safety, public security, fundamental rights, or the society as a whole, that can be propagated at scale across the value chain}" (Article 3, (44d)) \cite{aiact}.
    
    Some examples of high-risk systems are the one that operates in the educational and professional sphere (i.e. admission to university and job hiring), in access to essential private services and essential public services or administration of justice and democratic processes such as systems used by a judicial authority to assist it in researching and interpreting facts and law. 

    In this paper, we focus on this type of decisions, that are generally binary with a "good" and a "bad" outcome. Ensuring fairness in these decisions is essential as they can have unintended consequences on people's lives and can deeply influence the future trajectory or stability of individuals, or even communities.

    In addressing concerns about fairness in high-stakes decisions, we must consider what stage of the decision-making process we are evaluating for fairness. Is it the fairness of the process itself, the recommendations provided, or the final decisions made? As discussed in \ref{section2_1}, many ADMS actually "suggest" a certain action, which is generally followed by the decision-maker, as observed in credit scoring and predictive justice scores. Therefore, when discussing fair outcomes in this context, it is primarily about fair recommendations.
    
    Fair outcomes involves the absence of bias or discrimination and should be considerate of the interests of all stakeholders.
    However, it's important to note that there is no universal definition of fairness it depends on the notion adopted by the decision-maker, as will be further discussed in the next section.
    
    Fairness of the process, on the other hand, requires that the decision-making process should be explainable, justifiable and perceived as meaningful for the analyst, useful for the decision-maker and legitimated by stakeholders \cite{tsoukias2007concept}. The process might need to be understandable enough to be argued by the stakeholders and possibly be challenged or recused. For this to be possible, we consider that providing explanations for which the automated system made a given recommendation is fundamental and a first step to ensure fairness \cite{tsoukias2020social}.

    In this paper, the proposed approach is positioned within the framework of process fairness in high-stakes decisions. However, this should not be interpreted as underestimating the importance of fairness of recommendations. 
    A lot of research in social psychology explored the link between fair process and fair outcome, notably the work of Lind et al. (see \cite{lind1988social, van1998evaluating, lind1990voice}). They have demonstrated that fair processes can significantly influence how individuals react to outcomes. For instance, participants who were afforded the opportunity to express their opinions during the process tended to react more positively to the outcome compared to those who were not given this opportunity \cite{van1998evaluating}. This remains out of the scope of this paper, we will only focus here on process fairness.

    \subsection{Fairness for whom ?}
    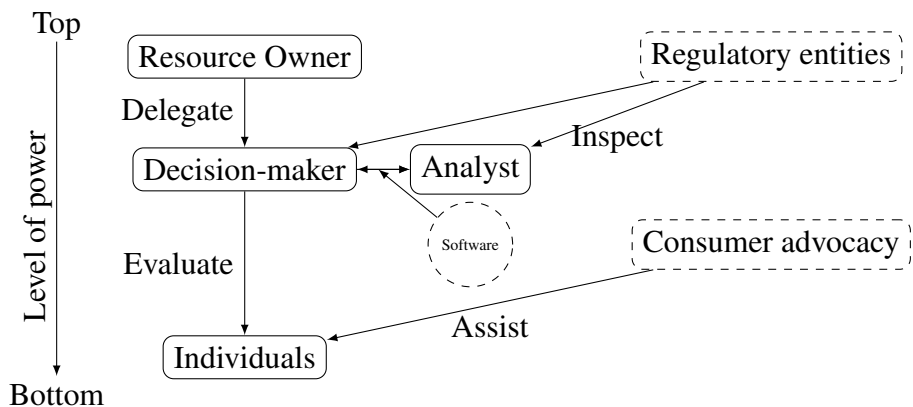
\begin{figure}[ht]
        \centering
        \begin{tikzpicture}[node distance=1.5cm and 7cm, >=latex, scale=0.5]
    \node (owner) [rectangle, draw, rounded corners] {Resource Owner};
    \node (custodian) [rectangle, draw, rounded corners, below of=owner, align=center] {Decision-maker};
    \node (experts) [rectangle, draw, rounded corners, right of=custodian, node distance=3cm] {Analyst};
    \node (regulator) [rectangle, draw, rounded corners, dashed, right of=owner, node distance=7cm] {Regulatory entities};
    \node (individuals) [rectangle, draw, rounded corners, below of=custodian, node distance=2.5cm] {Individuals};
    \node (software) [circle, draw, dashed, below of=experts, node distance=1cm] {\tiny Software};
    \node (advocacy) [rectangle, draw, rounded corners, dashed, right of=software, node distance=4cm] {Consumer advocacy};
    
    \draw[->] (owner) -- node[midway,left] {Delegate} (custodian);
    \draw[<->] (custodian) -- node[midway,above] {} (experts);
    \draw[->] (regulator) -- node[midway,below] {Inspect} (experts);
    \draw[->] (regulator) -- (custodian);
    \draw[->] (custodian) -- node[midway,left] {Evaluate} (individuals);
    \draw[->] (advocacy) -- node[midway,below] {Assist} (individuals);
    
    \draw[->] (-5,0.4) -- (-5,-8.5) node[rotate=90] at (-5.5,-4.5) {Level of power};
    \node at (-5,0.8) {Top};
    \node at (-5,-9) {Bottom};
    
    \draw[->] (custodian) -- (experts);
    \draw[->] (software) -- (3.55,-3);

\end{tikzpicture}
        \caption{Stakeholders in a decision-making process}
        \label{fig:stakeholders}
    \end{figure}

    We refine the framework to clarify to whom we address fairness in the decision process. As discussed previously, we recognize the involvement of multiple stakeholders in a DP. 
    In figure \ref{fig:stakeholders}, we illustrate one possible configuration of stakeholders, in the case of high-stake decisions, and expose the power dynamics between them. 

    Power can be defined in various ways: according to Crozier \cite{crozier1990pouvoir}, it's a dynamic relationship where one party can gain more than the other, yet neither is entirely destitute against the other. Dahl \cite{dahl1971gouverne}, on the other hand, views power as the ability of one party to ensure favorable terms of exchange for him/her in negotiations with the other.

    Exploring the foundations of power, as suggested by Crozier \cite{crozier1990pouvoir}, it's rooted in the assets, resources, and strengths of each stakeholder. However, it's not limited to these factors; the ability to take action (\textit{possibilité d'action}) is also crucial. When A holds power over B, it's not merely to exert control but because A seeks to achieve an objective and B influences the possibility of its attainment. Therefore, power also lies in the \textit{level of freedom} each stakeholder possesses and their ability to decline demands from others. Additionally, factors such as authority and other subjective attributes significantly impact freedom of action.

    Adopting this paradigm, we outline the stakeholders involved in the decision-making process and position them along a power axis (depicted in Figure \ref{fig:stakeholders}) to better understand the dynamics of their relationships and how do we want to move along this axis to approach the decision-making process differently. 
    
    At the top, we place the resource owner, representing the actor(s) with ownership rights over the resource to be distributed. They may possess financial, educational, or other forms of capital. It's important to note that the presence of this stakeholder is contingent on the case involving quantifiable resources controlled by individuals. For example, in the context of credit lending, the resource owner would be the investors and shareholders.

    Below the resource owner, we find the decision-maker, tasked with allocating the resource based on interactions with analysts possessing technical knowledge to propose models (i.e. a software) and conduct proper evaluations for decision-making. At the lowest tier are the individuals impacted by the decision, often lacking influence or agency in the decision-making process. We specify "often" because in some cases users have the possibility of recourse to contest a decision and win the case, notably with the help of another eventual stakeholder who would be the consumer advocacy that help to assist citizens to claim their rights. 
    
    Additionally, decision actors are bound by a fiduciary duty to legal principals responsible for ensuring compliance with relevant legislation. Indeed, a decision goes through a legitimization process. What partially confer legitimacy to decisions made by one of the stakeholders are the established norms, rules and objectives. Indeed, actions are assessed against norms, results are compared to objectives, procedures are judged compliant with reference to the rules \cite{claude1996pouvoir}. 

    It is important to note that fairness is always established with respect to a specific stakeholder. We can consider that a process is fair if it is fair towards all stakeholders. However, achievement of fairness can sometimes be conflicting. For example, in credit allocation, "positive" recommendations might be considered fair for socially deprived individuals who may not be financially solvable. Granting credit to these individuals can be viewed as an act of social justice. However, this can also be seen as unfair to the resource owner because it jeopardizes the financial stability of the banking institution.
    In this framework, we center our concerns about fairness towards the individuals impacted by the decision. Even though, fairness towards other stakeholders shouldn't be overlooked, the trade-off between fairness and performance will not be treated in this paper.

\section{Critics towards traditional algorithmic fairness}
\label{section3}
\subsection{Group fairness limitations}
Some of the first work on fairness in the literature often center around what we call \textit{group fairness} \cite{hardt2016equality, feldman2015certifying, zafar2017fairness, verma2018fairness}. This concept originates from the recognition that certain sub-populations have endured historical discrimination in our society and that was embedded in the data that we use to train models nowadays. Consequently, the goal of group fairness is to rectify biases by considering a protected attribute (e.g. gender, race, nationality), creating groups based on attribute modalities and mitigating outcomes within those groups. 

For example, let's consider an employer that relies on a decision-aiding process, where given an application (i.e. a vector of individuals' characteristics) returns a prediction about whether a candidate would make a good employee. Applying group fairness in this process could be a way to ensure equal opportunities for all genders, acknowledging and correcting past gender biases.  

Despite its intent, group fairness faces certain limitations when confronted with the complexity of historical biases and their translation into algorithmic decision-making. In the following, we set out the shortcomings we have observed.

\subsubsection{Inherent incompatibility}
There is an inherent incompatibility between some fairness metrics. This limitation has been extensively studied by \cite{kleinberg2016inherent}. While we won't focus on this aspect, it is worth mentioning their impossibility theorem that highlights the tension between different definitions of fairness, such as equalized odds, equal opportunity, and calibration, when applied to risk prediction. The authors demonstrate that in certain scenarios and given a protected attribute, achieving one form of fairness may inevitably lead to unfairness in another dimension.

\subsubsection{Inner-group inequalities}
Group fairness approaches often ignore inner-group inequalities which refer to disparities or variations that exist within a particular demographic group. Indeed, discrimination is blind to the definition of groups so it is important to consider intersectionality, which acknowledges that individuals have multiple dimensions of identity. 

The concept of intersectionality, introduced by Crenshaw \cite{crenshaw1991mapping}, highlights how overlapping systems of power impact the most marginalized individuals. While rooted in the analyses of black feminists, intersectionality has been embraced by various sociological perspectives, including Marxism, Weberianism, and Bourdieusian sociology \cite{yuval2015situated}. It challenges traditional stratification theory by rejecting a singular focus on one social division (i.e. class), recognizing that differentiation occurs across various facets of social analysis. Individuals are actually positioned along socioeconomic grids of power, identificatory perspectives of belonging, and the normative value systems that shape their experiences within a complex societal framework \cite{yuval2015situated}. This brief overview lays the groundwork for deeper exploration in future research.

Therefore, it appears unrealistic and unjustifiable to center the evaluation of a process's fairness solely on a single protected attribute. Such an approach is reductionist, oversimplifying people's identities. Even if fairness is attained within one dimension, it disregards other forms of discrimination. For instance, imagine that an employment interview process achieves statistical parity between men and women, meaning both groups receive an equal number of interviews. According to this metric, the process appears fair. However, this approach does not detect the fact that black women receive significantly fewer interviews than white women, leading to unfair practices. 

Even though some have suggested constructing subgroups with various combinations of protected attributes \cite{hebert2018multicalibration}, the challenge lies in their large number, making it difficult to consider all possibilities.

%--------------------------------------------------------------
\subsubsection{On the legitimacy of discrimination based on "merit"}
The tension between maximizing an objective and predictive parity as a fairness metric has been discussed in economics as a failure of predictive parity to reflect taste-based discrimination \cite{kasy2021fairness}. In general, fairness metrics rely on the notion of merit, which can inadvertently legitimize and perpetuate discriminatory practices aligned with the decision-maker's objectives. Usually, concerns for fairness arise when societal ideals are in tension with a decision-maker’s interests. The objective function's choice is crucial to the model. Most of the time it is chosen by policymakers who have ownership and control rights over the data and the algorithm \cite{CorbettDaviesetal2017}. Consequently, such metrics when applied on the objective function, may inadvertently reinforce the legitimacy of existing norms \cite{kasy2021fairness}. 

Usually, when we rely on merit to assess the entitlement of individuals to resources in an allocation process, we overlook the challenge of accurately measuring merit. The feature space employed often serves as a representation of latent or unmeasurable constructs. Friedler et al. \cite{friedler2021possibility} shed light on a compelling example in the context of college admissions: the decision should be made based on characteristics such as the intelligence, perseverance and motivation of candidates which are not directly observed. Rather, the feature space is constructed using proxies to those qualities such as IQ, school grades, extracurricular activities... Yet, these approximations may suffer from structural bias due to social and economic circumstances resulting in an incomplete quantification of the candidates' intrinsic characteristics.

If we try to analyse this from a social science perspective, we notice that defining merit is similar to understanding philosophical concepts like value and worth \cite{madan2007sociologising}. Habermas \cite{habermas1990moral} highlighted that issues once thought purely philosophical now require consideration of their social context. Philosophy must recognize its connection to real-world contexts and history \cite{honneth1991communicative}. Following this principle, we take interest in sociology, especially the sociology of education, where there has been thorough examination of the nature and constraints of merit. 

Bourdieu \cite{bourdieu1971reproduction} showed how the educational system reproduces the class structure, reinforcing social inequalities and social stratification, and conceals the fact that it fulfils this function under the guise of neutrality. In fact, the reproduction of these hierarchies, founded on the hierarchy of donation (''dons'' in french) and merit, serves as a legitimization for the perpetuation of the social order.
However, the cultural capital and the capital of relationships inherited from the family ''\textit{are the condition, if not the main factor, for success}'' \cite{bourdieu1971reproduction}. Therefore, the relative autonomy of the educational market only lends apparent justification to the meritocratic ideology. However, this overlooks that intelligence or academic diligence represent only one form of capital, often possessed alongside economic and social capital. Moreover, those with economic capital have greater chances of acquiring cultural capital, rendering educational credentials valuable primarily within the confines of the educational market \cite{bourdieu1971reproduction}. 

%--------------------------------------------------------------

\subsubsection{How about fairness in other fields ?}
In other disciplines besides the computer science literature, fairness exists independently of the existence of a protected group. Actually, there is no or little connection between how fairness is treated in the computer science community and the notion of fairness in social choice theory or welfare economics.

In social choice theory, the focus is on individual fairness through an axiomatic approach and the purpose is to distribute resources or make collective decisions in a fair manner by aggregating individual preferences. One of the principles of fair division in social choice (\cite{brams1996fair, moulin2004fair, brandt2016handbook}) is envy-freeness: each person should receive a share that is, in their eyes, at least as good as the share received by any other agent. In other words, an allocation is considered envy-free if no agent would prefer someone else's bundle over their own. 
Contributions that study fair classification from fair division perspective are discussed in the related work section \ref{related}.

In welfare economics, Sen \cite{sen1979equality} extends Rawls' conception of primary goods to focus on what goods actually do for individuals. This shift forms the basis of the concept of 'capability equality'. 
He argues that an individual's well-being cannot be adequately measured solely by the resources they possess, but must also include their capability to use these resources to live a life they have reasons to value. This perspective ensures that the assessment of well-being and fairness includes the real opportunities available to individuals, reflecting their own perceptions and values. 

Ed Diener has enriched the literature on psychology with a subjective well-being (SWB) model \cite{diener1984subjective}. He showed that objective conditions such as health and wealth are not inherent and necessary to evaluate the SWB of people. 
Instead, SWB describes how people perceive the quality of their lives. This led to a proliferation of welfare economic research using subjective measures of happiness and life satisfaction \cite{kahneman2004toward}. 
It's worth mentioning that an individual's perception of fairness holds significance for their sense of well-being and can indeed impact social interactions, thus contributing to the overall well-being of society.

Since we try here to adopt an interdisciplinary perspective on fairness and explore insights from other fields, the subjectivity comes as an area worth exploring. The preferences of individuals should play a role in determining the fairness of a process. 

%--------------------------------------------------------------

\subsubsection{Non-universality}

The categorization between protected and unprotected groups is often rooted in political constructs rather than scientifically grounded boundaries, so it presents challenges when we try to universally generalize this approach. The term "protected" usually refers to protection from anti-discrimination laws or policies notably with respect to gender, "race", ethnicity, religion, sexual orientation etc..
We note that the outdated notion of race as a scientific concept, has been widely discredited and is now recognized as entirely erroneous. Here we are referring to the sociological concept of "social race" as the socially constructed racial categories.
Some of the political and social factors that shape how these categories are formed and recognized are societal perceptions, historical contexts, power dynamics, legal systems and cultural norms.

For instance, in the United States, racial categories in census are commonplace. The country has a long history of racial discrimination, redlining and partisan gerrymandering. During the eighteenth century, those in positions of political power perceived race as an inherent and obvious aspect of human identity, aligning with the ideals of the European Enlightenment. Ever since, race was an organising ideology and census of the population integrated racial categories, which varied through time \cite{nobles2002racial}. However, after the civil rights movement, the purpose of racial categorization drastically changed. These information are now used to establish public policies that respect civil rights and to dismantle discriminatory mechanisms as residential discrimination, exclusion from certain occupation etc... \cite{nobles2002racial}. Hence, in this context, racial data seem to be a good way to evaluate the fairness of government programs and to monitor compliance with anti-discrimination laws and regulations \cite{kertzer2002census}.
  
In parallel, in the majority of the EU member states, notably France and Germany, the distinction based on race is nonexistent. Collecting data on racial and ethnic origin is actually prohibited pursuant to the Racial Equality Directive (RED), which was adopted in 2000.  
It is an important legal instrument within the EU aimed at eliminating discrimination and ensuring equal treatment and opportunities for all individuals, irrespective of their racial and ethnic differences. This aligns closely with the concept of "fairness through blindness" found in the scientific literature. 

However, the directive has led to a unique approach in these countries, where the concept of race isn't officially recognized or used for data collection purposes. This approach has its roots in historical sensitivities and experiences, particularly in countries like Germany, where the legacy of Nazi ideology and its catastrophic consequences have heavily influenced national policies regarding racial categorization \cite{chin2010after}.

France, similarly, has historically followed a republican model of "colorblindness" or "universalism," aiming to promote equality by avoiding the recognition of racial or ethnic distinctions within its legal frameworks. This stance is rooted in the idea that acknowledging such differences might lead to division and inequality \cite{chapman2004race}.

Therefore, implementing a universal approach based on considering protected attributes poses significant challenges. While this method might hold relevance within the context of the historic and political situation in the USA, its applicability elsewhere is not straightforward.

%--------------------------------------------------------------

\subsubsection{Do not assume sense of belonging !}
Constructing demographic groups is generally done by the decision-maker (or data collector) based on assuming and inferring the affiliations. 
The construction of demographic groups typically relies on the decisions made by those in positions of power, often relying on assumptions and perceived characteristics about individuals' identities. Thus, this process of assumption-making can be inherently flawed or oversimplified, and can perpetuate stereotypes and biases, potentially leading to discrimination or unequal treatment based on these constructed demographic groups.

Indeed, social identification and belonging to a community can be a matter of choice.  An empirical study \cite{obst2007choosing} aimed at investigating the relationship between the degree of choice in community membership and the subsequent levels of social identification  showed that higher degree of choice is associated with higher levels of cohesiveness within a community.  Membership to a community of interest may even be stronger than identification to a local neighborhood for example.
Hence, deducing community affiliation becomes unsatisfactory due to the potential for individuals to opt for alternative group memberships beyond their designated assignment. For instance, a person classified as male in their civil status may choose not to align themselves with the male category and identify with a different gender group. 

The choice of belonging to a community is political and not a natural process. 
Taking the instance of race, it is indisputable that it emerges as an ideological construct. The illustration of the multiracial scenario amplifies this claim as individuals may belong to multiple racial categories. The imposition of inevitably invented racial categorizations is often resisted, as it contradicts self-perceptions and undermines the authenticity of personal feelings.

%--------------------------------------------------------------

\subsection{Individual fairness limitations}

In contrast to group-based fairness, some works explored individual fairness according to the principle \textit{"equals should be treated equally"} \cite{dwork2012fairness}. This notion aligns more closely with our definition of fairness, recognizing that the assessment of fairness should occur at an individual scale.
Building on this approach, we identify two limitations that we aim to overcome.

First, this principle relies on the notion of "equals". But what can we consider "equals" ? In the literature, equality is measured through similarity on objective criteria such as income, wealth, education... It aims to ensure that individuals have the same opportunities and access to resources, if they are sufficiently similar on those objective criteria regardless of their background or (irrelevant) personal characteristics. However, this makes us question the legitimacy of those objective criteria and who has the right to fix them. 
We also point out that this principle fails to capture the subjectivity of equality, as "equals" is very different from "feeling equal". The former is a judgment held by the decision-maker while the latter takes into account the perceptions of the individuals impacted by the decision. 
Indeed, "subjective equality" focuses on how individuals experience equality, taking into account people's feelings and perceptions of fairness and justice. It recognizes that equality isn't solely about objective measures but also about how individuals perceive their treatment in society. \\
For example, two individuals with the same objective level of income may still feel unequal if one perceives their income as unfairly low due to discrimination or systemic barriers. Conversely, two individuals living in distinct environments with significantly different incomes may still perceive themselves as equal if their purchasing power remains similar.

Second, defining similarity between individuals is not an easy task and it has been one of the challenges of Dwork's work \cite{dwork2012fairness}. However, it has generally been assumed that this distance metric would be symmetric.
To our knowledge, no approach in the fairness literature quantifies similarity using non-symmetric distance measures. Yet, this constitutes a crucial aspect of our work. Indeed, we believe that there is a non-negligible subjective dimension in the notion of similar (or equal) that must be reflected in the chosen similarity measure. Having a symmetric distance constrains the subjectivity of similarity, as we will always rely either on objective features or on one of the two parties to determine a real-value distance. Therefore, if we aim to quantify the similarity between two individuals, we must consider that they may not perceive their distance in the same way. We can get a sense of why this is important through an example. 

Let's recall the property of individual fairness according to \cite{dwork2012fairness}: a mapping $M$ satisfies individual fairness (IF) if for all $(x,y) \in X^2$, we have $D(M(x), M(y)) \leq d(x,y)$. 
In this example, let $M$ being used for college admission decisions, yielding a probability that an applicant should be admitted. Suppose two applicants $x$ and $y$ that are objectively quite similar but $y$ has slightly higher scores at exams so the objective distance between the two is $d(x,y)=0.05$. And the mapping produces these scores $M(x)=0.85$ and $M(y)=0.9$. Then, the property $D(M(x), M(y)) \leq d(x,y)$ holds and $M$ would be considered fair. However, $x$ may perceive that she is closer to $y$ because despite coming from a more disadvantaged high school and having less guidance and resources, $x$ has exerted more effort than $y$ to achieve similar scores, so $x$ believes that $d_x(x,y)=0.04$. Consequently, $D(M(x), M(y)) > d(x,y)$, leading $x$ to perceive the treatment as unfair, while $y$ adheres to the objective distance measure ($d_y(x,y)=0.05$) and views the outcome as individually fair.
The use of non-symmetric distance function in this example enables us to incorporate the perceptions of the individuals, which is the focus of this paper.

\section{Subjective fairness}
\label{section5}
    Building on the analysis of fairness measures' limitations, we try to redefine fairness with a subjective dimension, taking into consideration the perception of individuals. 

    One way to conceptualize this shift is by transitioning from a top-down to a bottom-up approach. When it comes to determining what constitutes equality, the responsibility should not rest exclusively with decision-makers. Instead, impacted individuals should be empowered to identify and report potential instances of discrimination, and have a say in evaluating whether they perceive their treatment as fair \cite{gentelet2021strategies}. 
    Our definition of fairness can be roughly summarized as a subjective extension of "equals should be treated equally". 
    
    Let's consider a set of individuals $\mathbf{I}$, a set of issues $\mathcal{X}$ and a set of outcomes $\mathbf{R}$.
    A decision-support system is a mapping $M : \mathbf{I} \times \mathcal{X} \mapsto \mathbf{R}$ such that $M(x,\psi) = r \in \mathbf{R}$ refers to the outcome $r$ where individual $x \in \mathbf{I}$ is concerned by issue $\psi \in \mathcal{X}$.
    
     We introduce a non-symmetric subjective similarity measure such that given an individual $x \in \mathbf{I} \text{, } \forall z \in \mathbf{I} \; sim_x : \mathbf{I}^2 \mapsto [0,1]$ describes the similarity between two individuals in the set $\mathbf{I}$ as perceived by $x$. 
    From that we construct the set $S_x$ of individuals that $x$ considers similar to him such that
    \begin{equation*}
        S_x = \{z | z \in \mathbf{I}, sim_x(x,z) \geq \delta\}
        \label{S_i}
    \end{equation*}
    
   We then assume a similarity metric $ T : \mathbf{R}^2 \mapsto [0,1]$ describing the similarity between the treatment of individuals, and define $\phi (x, \psi)$ being the fact that $x$ considers being treated fairly on purpose $\psi$. It could be binary $\phi(x,\psi) \in \{\textit{fair}, \textit{unfair}\}$ or take the form of a scale, a score, etc.
    To initiate this process, we put forth an initial definition of subjective fairness on an individual scale.
    
    \begin{definition}[Individual Subjective Fairness (ISF)]
        Given $x \in \mathbf{I}$ and $\psi \in \mathcal{X}$, $x$ considers herself to be treated fairly with respect to $\psi$ if all individuals she considers similar to herself are treated similarly : 
        \label{subjfairness} 
        \begin{equation}
        \phi_{\delta, \epsilon} (x, \psi ) \Leftrightarrow \forall y \in S_x \text{, } T(M(x,\psi), M(y,\psi)) > \epsilon
    \end{equation}
    \end{definition}

    \begin{definition}[Subjective fair process]
        A decision process is $(\delta, \epsilon)-SF$ if all individuals involved in the decision process consider themselves being treated fairly: 
        \begin{equation}
            F_{\delta, \epsilon} (\mathbf{I},\psi )  \Leftrightarrow \forall x \in \mathbf{I} \text{, } \phi_{\delta, \epsilon} (x, \psi ) = \textit{fair}
        \end{equation}
    \end{definition}

    \begin{example}
         Let's consider two employees, \textit{Alice} and \textit{Bob}, both applying for a raise. Objectively, they have similar qualifications. For instance, both hold master's degrees and possess relevant job experience. However, their social status differ significantly.
        \textit{Alice} is a black women, who comes from a privileged background, attended a prestigious school, and had financial support throughout her education.
        \textit{Bob} is a white man, who comes from a precarious environment, had to work part-time jobs to pay for school, and attended a less renowned institution.
        
        Suppose Alice considers herself similar to Bob due to their shared qualifications. If Bob is accepted while Alice is not, it suggests disparate treatment. Alice may perceive the salary increase process as potentially sexist and/or racially discriminatory, given that the only discernible difference between her and Bob is their social backgrounds.

        Now, let's suppose that Bob does not consider himself as similar to Alice. In such a case, Alice's outcome has no impact on Bob's perception of fairness. Since he doesn't view her as part of his group, he doesn't anticipate receiving the same treatment as her.
 
        \label{def32}
    \end{example}

    As we are dealing with subjective information, including feelings and opinions, we are confronted with defeasible information. For instance, perceptions of similarity may fluctuate due to evolving factors, leading to shifts in their understanding or evaluation of a given scenario.
    
    Consequently, the determination of subjective fairness is not a static, one-time event but rather a dynamic process that necessitates justification for stakeholders to evaluate their perceptions effectively. In the following, we explore this point further. 
    
\section{Explanations to achieve subjective fairness}
    The issue of explainability is not independent of fairness. We believe that fairness could be achieved through explanations. By understanding the reasoning and the functioning of an algorithm, we gain the ability to identify instances of discrimination and rectify them, ensuring fairness towards all stakeholders. Multiple works operate between these two topics including \cite{dodge2019explaining, zhao2023fairness, begley2020explainability}.

    First, we can establish how explanations could justify the fairness of a process in accordance with a normative standard. This is commonly known as \textit{procedural fairness}. It adopts an objective stance, asserting that fairness is contingent upon adherence to procedural rules \cite{leventhal1980should, kaufmann1970legality}. Leventhal \cite{leventhal1980should} has posited that procedural fairness often hinges on satisfying six key constraints. Some of these principles are:
    \begin{itemize}
        \item[-] \textit{The consistency rule} states that the process should remain consistent and uniform across all individuals, aligning closely with the concept of equality of opportunity.
        \item[-] \textit{The accuracy rule} dictates that the decision process should rely on the best available information, reflecting the principle of accountability.
        \item[-] \textit{The ethicality rule} requires that procedures must be compatible with the fundamental moral and ethical values accepted by the stakeholders. 
    \end{itemize}  
    
    Implicit in this approach is the notion that fairness can be assessed at a specific moment, guided by predefined rules, overseen by the decision-maker, and remains static without evolution.
    
    \begin{definition}[Fairness through objective explanations]
        A process is fair if the explanations about how the process has been conducted either satisfy pre-established rules or a normative standard.
    \end{definition}
    
    Although fair procedures are commonly perceived as neutral and free from self-interested or ideological considerations, individuals make subjective evaluations of processes' fairness since it depends on their knowledge, prior preferences and bias.
    This is supported by \cite{doherty2012ends}, as they state that "\textit{The idea that perceptions of procedural justice are subjectively determined is demonstrated in that appraisals of what is fair and unfair can vary across both individuals and circumstances. [...] people discount the importance of fair procedures when they were motivated to find support for (or arguments against) the legitimacy of a given outcome.}"
    
     Objective explanations are particularly suitable to justify and support decisions for stakeholders not directly affected by them, but who are more interested in evaluating the process and ensuring regulatory compliance. However, in our framework for achieving subjective fairness (towards impacted individuals) through explanations, compliance with normative measures is necessary but not sufficient. Explanations should serve an additional role which is proving the legitimacy of the decision.

    Following this idea, we consider that explanations are a social interaction process between two parties. It builds upon descriptive explanations to present an argument that convinces the stakeholders of the fairness and legitimacy of a decision. 
    These arguments are defeasible, in the sense that they are not absolute and can be overturned when new information or a change in perspective is introduced. The level of acceptance of an argument is subjective and varies depending on the beliefs, attitudes, and biases of the audience. The same argument may be persuasive to some people but not to others \cite{bench2007argumentation}.
    
    \begin{definition}[Subjective fairness through explanations]
        A process is fair if the explanations about how the process has been conducted are convincing and accepted by all the population.
        \label{def6}
    \end{definition}

    \begin{example}
        In example \ref{def32} where Alice considers herself as similar to Bob, she expects that they will receive the same treatment. However, the employer decides to only grant a raise to Bob. 
        One plausible explanation for this decision could be that Bob brought in a substantial client to the firm. According to their employment contract, employees who secure new clients and sign service agreement exceeding a certain amount are eligible for a promotion (normative explanation). Plus, a justification advanced by the decision-maker is that the firm has limited resources, and the raise is directly tied to the value of the signed agreement. 
        If Alice accepts this argument, we can conclude that the process is subjectively fair for both Alice and Bob. Otherwise, Alice should advance counter-arguments to contest the decision. If there are no further explanations to respond to Alice's arguments, then the process is considered unfair.
    \end{example}

    \subsection{Explainable clustering models}
    The choice of the clustering model must prioritize explainability. There are several methods to achieve this. First,
    a priori clustering which includes unsupervised machine learning algorithms. While effective, these methods rely on correlations rather than causal relationships and often depend on objective features, which may not capture subjective aspects \cite{cliff1983some}. Incorporating causality into fairness frameworks has been suggested to address these limitations \cite{makhlouf2020survey}. Notably, semi-supervised learning frameworks that allow for the integration of user feedback to introduce subjectivity into the clustering process \cite{hoffmann2019fairness}.
    
    Fitted clustering is a method that involves collecting individual perceptions and preferences through surveys or interviews to create clusters reflecting shared subjective viewpoints. While theoretically robust and explainable, it is challenging to implement on a large scale and can lead to inconsistencies when integrating multiple perspectives. This situation is discussed in the next subsection.
    Similarly, sample-based clustering uses a sample of candidates to derive clusters. It is easier to implement but assumes the sample is representative of the entire population, which may not always be the case.
    
    Additionally, Raboun et al. \cite{raboun2023dynamic} introduced a dynamic clustering which addresses limitations in generating ratings by categorizing objects into predefined classes based on preference relations and reference profiles. It dynamically updates preferences with each new rating, ensuring explainability and consistency.

    \subsection{Decision on clusters}

    In the following, we propose a framework where we have recommendations of the decision-support system and the objective is to ensure that individuals who perceive themselves as similar receive the same treatment. Given subjective groups based on individuals' perceived characteristics, our focus is to make decisions that satisfy SF.
    
    Let $\mathbf{I}$ be the population of individuals with $|\mathbf{I}|=n$. 
    Each individual $i \in \mathbf{I}$ is associated with the set $S_i$ of individuals he considers similar to himself, as presented in equation \ref{S_i}.  We can denote the population as $X = \bigcup_{i \in \mathbf{I}} \{S_i\}$.

    \begin{assumption}
        Each individual naturally perceives themselves as similar to themselves: $\forall i \in \mathbf{I} \text{, } S_i \neq \emptyset \text{ because } sim_i(i,i)=1 \text{, thus }i \in S_i$. This implies that we have a decision for every individual in the population. 
    \end{assumption} 
    One individual can belong to more than one cluster. 
    For example, consider two individuals $x,y \in \mathbf{I}^2$. Naturally, $x \in S_x$. Additionally, \( y \) considers \( x \) to be similar to her, so \( x \) is also included in \( S_y \). However, \( y \) is not necessarily included in \( S_x \) because we use a non-symmetric similarity function.
    
    Let $\mathbf{D}_\psi$ be a vector representing decisions concerning $n$ individuals for a purpose $\psi$. This vector is $n$-dimensional and consists of binary elements, where $\mathbf{D}_\psi : \mathbf{I} \rightarrow \{0,1\}$. Each element $d_i$ in $\mathbf{D}_\psi = (d_1, \ldots, d_n)$ corresponds to the final decision rendered for individual $i$.
    
    Let $\mathbf{R}_\psi$ be a vector representing recommendations suggested by the decision-support system for $n$ individuals in the population, for a purpose $\psi$, $\mathbf{R}_\psi : \mathbf{I} \rightarrow \{0,1\}$. Each element $r_i$ in $\mathbf{R}_\psi= (r_1, \ldots, r_n)$ corresponds to the recommendation for individual $i$.
    Recommendations could also be scores. 

    The objective here is to define the decision set $\mathbf{D}_\psi$. It is important that the decision-making process remains explainable, as we need to justify decisions to individuals. 
    In the following discussion, we present a solution that serves as a starting point and should be further enriched with explanations. We outline various recommendations that can be constructed and specify the explanation requirements for each of them.

    \textbf{Decision from individuals to sets} 
    Let's consider that we have $n$ clusters ($S_i, \; \forall i \in \mathbf{I}$). Each individual in the clusters carries a recommendation suggested by the ADMS. It is very likely to have different recommendations within a same cluster. In this case, SF is not satisfied. 
    We propose a relaxed version of ISF where, rather than comparing $x$ with every other individual individually, we compare $x$ with the collective outcomes of the individuals in her cluster as a group. We can imagine that for $x$, we seek an outcome that is as similar as possible to the aggregated outcomes of individuals in her cluster.
    
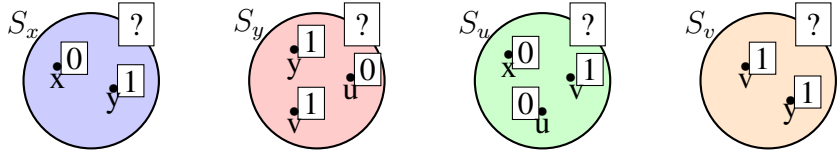
\begin{figure}
        \centering
    \begin{tikzpicture}[scale=1.5]
        % Cluster S_u
        \draw[thick, fill=blue!20] (0,0) circle (0.6);
        \node at (-0.6,0.5) {$S_x$}; 
        \node at (0.4,0.5) [rectangle, draw, fill=white] {?}; 
        \node at (-0.15,0.2) [rectangle, draw, fill=white, inner sep=2pt, outer sep=0pt] {0}; 
        \node at (-0.3,0.1) {\textbullet};
        \node at (-0.3,0) {x}; 
        \node at (0.35,0) [rectangle, draw, fill=white, inner sep=2pt, outer sep=0pt] {1};
        \node at (0.2,-0.1) {\textbullet};
        \node at (0.2,-0.2) {y}; 
       
        % Cluster S_y
        \draw[thick, fill=red!20] (2,0) circle (0.6);
        \node at (1.4,0.5) {$S_y$}; 
        \node at (2.4,0.5) [rectangle, draw, fill=white] {?};
        \node at (1.95,0.35) [rectangle, draw, fill=white, inner sep=2pt, outer sep=0pt] {1};
        \node at (1.8,0.25) {\textbullet};
        \node at (1.8,0.1) {y}; 
        \node at (2.45,0.1) [rectangle, draw, fill=white, inner sep=2pt, outer sep=0pt] {0};
        \node at (2.3,0) {\textbullet};
        \node at (2.3,-0.1) {u};
        \node at (1.95,-0.2) [rectangle, draw, fill=white, inner sep=2pt, outer sep=0pt] {1};
        \node at (1.8,-0.3) {\textbullet};
        \node at (1.8,-0.4) {v};
    
        % Cluster S_x
        \draw[thick, fill=green!20] (4,0) circle (0.6);
        \node at (3.4,0.5) {$S_u$}; 
        \node at (4.4,0.5) [rectangle, draw, fill=white] {?}; 
        \node at (3.7,0.2) {\textbullet};
        \node at (3.85,0.25) [rectangle, draw, fill=white, inner sep=2pt, outer sep=0pt] {0};
        \node at (3.7,0.1) {x}; 
        \node at (4,-0.3) {\textbullet};
        \node at (3.85,-0.2) [rectangle, draw, fill=white, inner sep=2pt, outer sep=0pt] {0};
        \node at (4,-0.4) {u}; 
        \node at (4.25,0) {\textbullet};
        \node at (4.42,0.1) [rectangle, draw, fill=white, inner sep=2pt, outer sep=0pt] {1};
        \node at (4.3,-0.1) {v}; 
    
        % Cluster S_y
        \draw[thick, fill=orange!20] (6,0) circle (0.6);
        \node at (5.4,0.5) {$S_v$}; 
        \node at (6.4,0.5) [rectangle, draw, fill=white] {?};
        \node at (5.8,0.1) {\textbullet};
        \node at (5.8,0) {v}; 
        \node at (5.95,0.2) [rectangle, draw, fill=white, inner sep=2pt, outer sep=0pt] {1};
        \node at (6.2,-0.2) {\textbullet};
        \node at (6.2,-0.3) {y}; 
        \node at (6.35,-0.1) [rectangle, draw, fill=white, inner sep=2pt, outer sep=0pt] {1};
    
    \end{tikzpicture}
    \captionsetup{font=small} 
    \caption{Example of recommended labels for individuals: some assignations are natural : $r_{S_v}=1$ whereas it is not clear what to assign to $S_y$, $S_u$ and $S_x$. According to majority voting $r_{S_x}=0$, $r_{S_y}=1$ and $r_{S_u}=0$}
    \label{fig:indiv2set}
    \end{figure}

    \begin{definition}[Relaxed ISF]
        Given an individual $x \in \mathbf{I}$ and purpose $\psi \in \mathcal{X}$, $x$ considers herself to be treated fairly with respect to $\psi$ if "most" individuals she considers similar to herself are treated similarly with respect to $\psi$ as she is treated: 
        \begin{equation}
            \phi'_{\delta, \epsilon} (x,\psi) \Leftrightarrow \forall y \in S_x \text{, } T(M(x,\psi), \text{agg}(M(y,\psi) \; \forall y \in S_x) > \epsilon
        \end{equation}
        \label{RISF} 
    \end{definition}
    \vspace{-0.9cm}
   One way to establish a single recommendation for every set $S_i$ is to aggregate the recommendations of all individuals in the set, as illustrated in figure \ref{fig:indiv2set}. Let $\mathbf{R}_{\text{set}}$ be a $n$-dimensional vector representing aggregated recommendations for each cluster $S_i, \; \forall i \in \mathbf{I}$, for a purpose $\psi$. The vector $\mathbf{R}_{\text{set}} : X \rightarrow \{0,1\}$. Each element $r_{S_i}$ in $\mathbf{R}_{\text{set}}= (r_{S_1}, \ldots, r_{S_n})$ corresponds to the recommendation for the set $S_i$, $r_{S_i}=1$ if a certain proportion of individuals in $S_i$ are labeled  $1$. It can be parameterized by $\theta$, if we want absolute majority, $\theta = \frac{1}{2}$
    
    \begin{equation*}
     \mathbf{R}_{\text{set}} \text{ is defined such as }r_{{S_i}}  = 
        \begin{cases} 
        1 & \text{if } \frac{1}{|S_i|} \sum_{x_j \in S_i} r_{x_j}  > \theta \\ 
        0 & \text{otherwise} 
        \end{cases}
    \end{equation*}  

    If each set $S_i$ has a unique label, we can encounter three scenarios:
    \vspace{-0.1cm}
    \begin{itemize}
        \item $T(r_{x},r_{S_{x}}) > \epsilon$ and $\forall y \in S_{x}, \; T(r_y,r_x) > \epsilon \Rightarrow$ ISF is satisfied
        \item $T(r_{x},r_{S_{x}}) > \epsilon$ and $\exists y \in S_{x}, \; T(r_y,r_x) \leq \epsilon \Rightarrow$ Relaxed ISF is satisfied
        \item $T(r_{x},r_{S_{x}}) \leq \epsilon \Rightarrow$ Neither ISF or relaxed ISF is satisfied
    \end{itemize}
    The recommendations $r_x$ proposed by the ADMS should be explainable. This involves providing explanations of the causal relationships between the subjective information provided by individuals and the final outcomes generated by the system. Additionally, the chosen aggregation method determines $r_{S_x}$ and thus should be justified. 
    These explanations enable to identify inaccuracies in the ADMS recommendations or erroneous or manipulative behavior of individuals. For example, explanations could give sufficient reasons that prove the dissimilarity between $x$ and the individuals in his cluster. Here after we present another way to make individual decisions by comparison to other clusters.
    
    \textbf{Decision from sets to individuals}
    Let's suppose we have a single recommendation for every set $S_i$, as computed above.
    Ideally, we could assign the outcome of the group to all individuals in that group: $\forall i \in S_x, d_i= r_{S_x}$. However, in reality, this process is not always straightforward for certain individuals who belong to multiple groups with conflicting outcomes. 
    
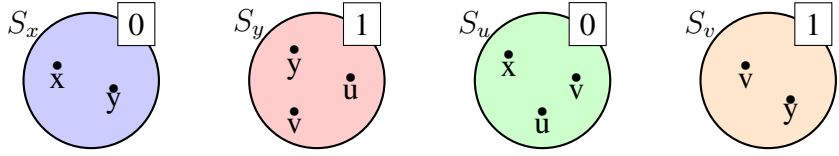
\begin{figure}
    \centering
    \begin{tikzpicture}[scale=1.5]
        % Cluster S_u
        \draw[thick, fill=blue!20] (0,0) circle (0.6);
        \node at (-0.6,0.5) {$S_x$}; 
        \node at (0.4,0.5) [rectangle, draw, fill=white] {0}; 
        \node at (-0.3,0.1) {\textbullet};
        \node at (-0.3,0) {x};
        \node at (0.2,-0.1) {\textbullet};
        \node at (0.2,-0.2) {y}; 
       
        % Cluster S_v
        \draw[thick, fill=red!20] (2,0) circle (0.6);
        \node at (1.4,0.5) {$S_y$}; 
        \node at (2.4,0.5) [rectangle, draw, fill=white] {1}; 
         \node at (1.8,0.25) {\textbullet};
        \node at (1.8,0.1) {y}; 
        \node at (2.3,0) {\textbullet};
        \node at (2.3,-0.1) {u};
        \node at (1.8,-0.3) {\textbullet};
        \node at (1.8,-0.4) {v};
        
        % Cluster S_x
        \draw[thick, fill=green!20] (4,0) circle (0.6);
        \node at (3.4,0.5) {$S_u$}; 
        \node at (4.4,0.5) [rectangle, draw, fill=white] {0};
        \node at (3.7,0.2) {\textbullet};
        \node at (3.7,0.1) {x}; 
        \node at (4,-0.3) {\textbullet};
        \node at (4,-0.4) {u}; 
        \node at (4.3,0) {\textbullet};
        \node at (4.3,-0.1) {v};
    
        % Cluster S_y
        \draw[thick, fill=orange!20] (6,0) circle (0.6);
        \node at (5.4,0.5) {$S_v$}; 
        \node at (6.4,0.5) [rectangle, draw, fill=white] {1}; 
        \node at (5.8,0.1) {\textbullet};
        \node at (5.8,0) {v}; 
        \node at (6.2,-0.2) {\textbullet};
        \node at (6.2,-0.3) {y}; 
    
    \end{tikzpicture}
    \captionsetup{font=small} 
    \caption{Example of recommended labels for clusters: some assignations are natural : $d_x=0$ whereas it is not clear what to assign to $y$, $u$ and $v$. According to majority voting $d_y=1$, $d_u=0$ and $d_v=1$ }
    \label{fig:clusters}
\end{figure}

    For instance, consider the case where $x \in S_x$ with $r_{S_x} = 1$, but also $x \in S_y$ where $r_{S_y} = 0$. How should we label $d_x$ in this situation? This scenario is illustrated in Figure \ref{fig:clusters}. Here after we reconstruct the set of decisions on each individual $i$ such as we aggregate all of the decisions that has been made on the sets in which $i$ belongs.
    The idea is to assign to $d_i$ the majority recommendation across all clusters that $i$ belongs to, eventually parameterized by $\theta$ as follow:

    \begin{equation*}
     \mathbf{D_\psi} \text{ is defined such as }d_{i}  = 
        \begin{cases} 
        1 & \frac{1}{|\{S_j \in X | i \in S_j \}|} \sum_{\{S_j \in X | i \in S_j \}} r_{S_j}   > \theta \\ %or put a parameter instead of 1/2
        0 & \text{otherwise} 
        \end{cases}
    \end{equation*}  
    
    As such, this method helps mitigate manipulative behavior to some extent. For instance, if individual $x$ strategically places themselves in a cluster with a higher likelihood of a favorable outcome, the fact that we take into consideration the recommendations of other clusters to which $x$ belongs helps neutralize $x$'s dishonesty. 
    This approach results in having three possibly conflicting outcomes: $r_x$, $r_{S_x}$ and $d_x$. When $r_x$ and $r_{S_x}$ are dissimilar we can either have:
    \vspace{-0.15cm}
    \begin{itemize}
        \item $T(r_{x},d_{x}) > \epsilon$: ISF criteria are not satisfied but could be justified with $d_x$ as we have to convince $x$ that they actually identify with a group to which there are enough reasons to be dissimilar to.
        \item $T(r_{x},d_{x}) \leq \epsilon$: ISF criteria are not satisfied, and the recommendation of $x$ is inconsistent with the recommendations of individuals in her cluster and individuals that put $x$ in their cluster. This may indicate that the ADMS has produced erroneous outcome and justifications about how this recommendation has been produced are required to clarify the decision that should be made. 
    \end{itemize}
    The explanations to treat conflicts have a subjective dimension, as they're tailored to individuals. We can talk about subjective explanations, but this will be further explored in future work.
    Beyond providing explanations, individuals should have the ability to engage in dialogue based on these explanations, making the decision process dynamic and responsive to both perspectives.

    \subsection{Discussion}
    The majority rule can pose limitations and potentially lead to unfair outcomes, especially when a majority of individuals can impose a great loss on a minority. Moreover, the majority rule is susceptible to inconsistencies and paradoxes, as highlighted by concepts such as the Condorcet paradox and Arrow's impossibility theorem \cite{brighouse2008democracy}. 
    
    Other forms of aggregation may be considered. For instance, proportionality involves aggregating individual outcomes based on their proportional contribution or significance within a group. This can be achieved by assigning weights to individuals based on specific criteria.
    For example, individual recommendation $r_x$ could have different weight according to the similarity between the recommendation of individual $x$ and others in her cluster $y$. As such, if $T(r_x, \text{agg}(r_y, \forall y \in S_x))$ is high, it means that $x$ accurately constructed their cluster. Consequently, we tend to trust $x$'s recommendation and give them higher weight if they are present in the clusters of the individuals in her group. 
    Another alternative could be a conflict resolution strategy that reconcile conflicting recommendations based on rules such as favoring the bad outcome over the good one.
    Also, we could look at individual objective attributes and specify rules or heuristics to resolve conflicts. For example, some rules could serve as a veto to a recommendation.

    \section{Related work}
    \label{related}
    %Related works of critics of group fairness
    Following the substantial increase in literature on fairness in artificial intelligence, numerous works have adopted a critical stance, advocating for an interdisciplinary perspective and emphasizing certain shortcomings in the approaches taken \cite{binns2020apparent, fleisher2021s, hoffmann2019fairness, john2022reality, golz2019paradoxes}. 
    In this vein, Binns \cite{binns2020apparent} argues that individual and group fairness are not inherently conflicting but rather represent different approaches to addressing the same moral and political concerns. They also contend that group fairness approaches may overlook discrimination stemming from intersectionality or groups not yet protected by anti-discrimination laws.
    Moreover, Fleisher \cite{fleisher2021s} highlighted the limitations of individual fairness notably the fact that the similarity measure is generally chosen by the decision-maker who holds their own bias and how moral values highly influence the choice of relevant features to determine similarity. 
    This work expands upon these ideas by integrating them with sociological insights, adding depth to the analysis.
    
    %Related works of subjective fairness
    Recognizing these limitations has driven some works to reconceptualize fairness beyond mere statistical metrics. Notably, Zafar et al. \cite{zafar2017parity} proposed a notion of fairness that is not based on parity (i.e. equality of outcomes or treatment) but on preferences. However, these preferences are related to sensitive attribute groups rather than individuals. Consequently, it ensures envy-freeness at the group level, guaranteeing that no group of users would be better off by changing their group membership. Therefore, this approach still operates at a group level, which is a concern we aim to avoid in this work.

    Concurrent work by Balcan et al. \cite{balcan2019envy} introduced an approach for fair classification tasks drawing from the literature of fair division, particularly using envy-freeness. It is adapted for scenarios with multiple potential outcomes, not just binary ones. Unlike the work of Dwork et al. \cite{dwork2012fairness}, which relies on a similarity function, this approach requires access to individuals' utility functions (preferences).
    The objective is similar to other machine learning approaches, aiming to minimize a loss function while satisfying some fairness constraint. Within this framework, the constraint is envy-freeness, which represents an individual measure of fairness that is based on the preferences of individuals rather than being a statistical measure chosen by the decision maker.\\
    Other works that connect fairness in decision-making with the literature of fair division in social choice include the Preference-Informed Individual Fairness (PIIF) framework of Kim et al. \cite{kim2019preference}. They introduce a relaxation of individual fairness (IF) and envy-freeness (EF), whereby the primary requirement is to satisfy IF, yet it remains flexible to accommodate individuals' preferences. However, it's worth noting that envy-freeness may not always be suitable for binary outcome problems. Conversely, our framework is situated within the context of high-stakes decisions where one outcome is universally perceived as "good" and the other as "bad", thus each individual will naturally prefer the "good" outcome.\\
    While our work shares the objective of eliminating envy, our approach diverges significantly, as we do not rely on individuals' preferences or their utilities over policies but use explanations to guide the perceived fairness of individuals. Indeed, individuals do not feel envy towards other if the explanations about how the decision has been made convince them of their fair treatment. The fairness of a process is established if all individuals accept both their own outcomes and those of others and is situated in a dynamic framework where evolving arguments can influence perceptions of similarity and acceptance.
    
    \section{Conclusion and future work}
        We introduced an innovative approach that extends beyond traditional objective measures of fairness by incorporating the subjective perceptions of individuals impacted by algorithmic decisions. This new framework aligns more closely with societal realities and empower individuals to determine their fair treatment.
        We then propose a novel definition of fairness and present our methodology that uses explanations as a tool. 

        As this paper is primarily conceptual on the notion of subjective fairness, future work will focus on a more comprehensive and rigorous conceptualization of explanations and justifications. Additionally, we will explore clustering methods and develop a robust explanation framework aimed at achieving subjective fairness.

        We also aim to address the challenge of treating conflicting fairness towards different stakeholders and managing conflicting explanations. Our objective is to take into account the perceptions of all stakeholders to ensure fairness for everyone. By incorporating diverse viewpoints, we develop a more inclusive and fair decision-making process that finds the trade-off between the varying interests of all parties involved.

        Furthermore, we plan to conduct experiments using a real dataset to evaluate the practical applicability and fairness of our approach in real-world scenarios.

%Bibliography
\bibliographystyle{splncs04}  
\bibliography{subj_fair}

\end{document}